**Oxygen-Octahedral-Tilting-Driven Topological Hall Effect in Ultrathin $SrRuO_3$ Films**

Youdi Gu[1,2,†], Yi-Wen Wei[3,†], Kun Xu[4,5,†], Hongrui Zhang[6], Fei Wang[1,7], Fan Li[2], Muhammad Shahrukh Saleem[2], Cui-Zu Chang[7], Jirong Sun[6], Cheng Song[2,*], Ji Feng[3,8,9,*], Xiaoyan Zhong[4,*], Wei Liu[1], Zhidong Zhang[1], Jing Zhu[4] & Feng Pan[2]

[1]Shenyang National Laboratory for Materials Science, Institute of Metal Research, University of Chinese Academy of Sciences, Chinese Academy of Sciences, Shenyang 110016, China

[2]Key Laboratory of Advanced Materials (MOE), School of Materials Science and Engineering, Tsinghua University, Beijing 100084, China

[3]International Center for Quantum Materials, School of Physics, Peking University, Beijing 100871, China

[4]National Center for Electron Microscopy in Beijing, Key Laboratory of Advanced Materials (MOE), The State Key Laboratory of New Ceramics and Fine Processing, School of Materials Science and Engineering, Tsinghua University, Beijing 100084, China

[5]Central Nano & Micro Mechanism, Beijing, Tsinghua University, Beijing 100084, China

[6]Beijing National Laboratory for Condensed Matter Physics, Institute of Physics, University of Chinese Academy of Sciences, Chinese Academy of Sciences, Beijing 100190, China

[7]Department of Physics, Pennsylvania State University, University Park, Pennsylvania 16802, USA

[8]Collaborative Innovation Center of Quantum Matter, Beijing 100871, China

[9]CAS Center for Excellence in Topological Quantum Computation, University of Chinese Academy of Sciences, Beijing 100190, China

†These authors contributed equally to this work.

*Corresponding authors: songcheng@mail.tsinghua.edu.cn (C.S.),

jfeng11@pku.edu.cn (J.F.) and xyzhong@mail.tsinghua.edu.cn (X.Y.Z.)

**Abstract**

Topological spin textures as an emerging class of topological matter offer a medium for information storage and processing. The recently discovered topological Hall effect (THE) is considered as a fingerprint for electrically probing non-trivial spin-textures. But the origin of THE in oxides has remained elusive. Here we report an observation of the THE in ultrathin ($\leq 8$ unit cells. u.c.) 4*d* ferromagnetic $SrRuO_3$ films grown on $SrTiO_3$(001) substrates, which can be attributed to the chiral ordering of spin structure (i.e., skyrmion-like) in the single $SrRuO_3$ layer without contacting 5*d* oxide $SrIrO_3$ layer. It is revealed that the $RuO_6$ octahedral tilting induced by local orthorhombic-to-tetragonal structural phase transition exists across the $SrRuO_3$/$SrTiO_3$ interface, which naturally breaks the inversion symmetry. Our theoretical calculations demonstrate that the Dzyaloshinskii-Moriya (DM) interaction arises owing to the broken inversion symmetry and strong spin-orbit interaction of 4*d* $SrRuO_3$. This DM interaction can stabilize the Néel-type magnetic skyrmions, which in turn accounts for the observed THE in transport. The $RuO_6$ octahedral tilting-induced DM interaction provides a pathway toward the electrical control of the topological spin textures and resultant THE, which is confirmed both experimentally and theoretically. Besides the fundamental significance, the **understanding of THE in oxides and its electrical manipulation presented in this work could advance the low power cost topological electronic and spintronic applications.**

**Introduction**

Topologically nontrivial spin textures have attracted extensive attention since they harbor elegant physics associated with the real space Berry curvature[1–3], and novel quantum transport phenomena and hold potential for energy efficient spintronic applications[4–6]. One intriguing example is the skyrmion spin texture, a topologically protected vortex-like object with swirling spin configuration. The magnetic skyrmion has been observed in B20 metallic magnets[7–9], interfacial asymmetric magnetic multilayers[10–12], and ferromagnetic oxide films, such as $(La,Sr)MnO_3$[13] and $SrRuO_3/SrIrO_3$ heterostructures[14,15]. When an electrical current flows through a skyrmion texture, the electron spin couples adiabatically with the skyrmion spin and experiences a fictitious magnetic field (i.e., Berry curvature) in real space. This field can generate an additional Hall conductance, which is a novel quantum phenomenon referred to as topological Hall effect (THE)[16,17]. The presence of THE is considered as an electrical transport signature of the skyrmion spin textures[5,14–22].

The formation of skyrmion in a thin film system needs the non-collinear Dzyaloshinskii-Moriya (DM) interaction expressed as $H_{DM} = \sum_{i,j} D_{ij} \cdot (S_i \times S_j)$ [12,23,24], where the $D_{ij}$ is the DM vector that determines the strength and sign of the DM interaction between a pair of spins $S_i$ and $S_j$. The DM interaction can be realized in a thin film system if the following two conditions are satisfied: (i) broken inversion symmetry, (ii) strong spin-orbit coupling (SOC). For example, the DM interaction can be generated in heavy metal/ferromagnet/oxide sandwiches[11,12,25], where the two interfaces and the heavy metal contribute to the broken inversion symmetry and SOC, respectively. Experimentally, a combination of the strong SOC of the ferromagnetic layer (e.g., CrTe and Mn-doped $Bi_2Te_3$) and broken interfacial inversion symmetry for single-layered CrTe[26] and Mn-doped $Bi_2Te_3$[27] thin films grown on $SrTiO_3$(111) substrates can also give rise to substantial DM interaction and concomitant THE. Recently, THE was observed in $SrRuO_3$/$SrIrO_3$ bilayers[14,15] and was explained by the formation of DM interaction stemming from the strong SOC of 5$d$ oxide $SrIrO_3$ and broken inversion symmetry in the bilayers. A natural and important question is whether there is a simplified approach to magnetic skyrmion and THE in $SrRuO_3$ thin film without the assistance of 5$d$ elements, especially since SOC is already strong in Ru.

Octahedral distortions such as deformation and tilting (or rotation) in oxides are currently of great interest, which provides broad opportunities for modulating physical properties of the oxide films, such as magnetic anisotropy, conductivity and exchange coupling[28–32]. The octahedral distortions can naturally break the inversion symmetry, thus provide a potential way to generate skyrmion spin texture. $SrRuO_3$, a rare example of 4$d$ band metal with a pseudo-cubic perovskite crystal structure, is a well-known itinerant ferromagnet with a Curie temperature ($T_C$) of ~160 K[33]. A number of intriguing electromagnetic phenomena have been demonstrated to be closely related to the Berry curvature in $SrRuO_3$, including magnetic monopoles[34], Weyl fermions[35], and even quantum anomalous Hall state[36]. These results indicate that the 4$d$ metallic ferromagnet $SrRuO_3$ should possess considerable spin-orbit interaction, as anticipated. Considering the broken inversion symmetry induced by octahedral tilting in the $SrRuO_3$ films, a non-vanishing DM interaction and resultant THE are highly expected in single-layered SrRuO$_3$ films. Moreover, since electric-field control of interfacial magnetism is of special interest for high-density and low-power consumption information storage[37], it is of great importance for electronic and spintronic applications to achieve an electrically tunable DM interaction. However, works that utilize an electric-field to effectively manipulate the DM interaction are scarce for oxide heterostructures. Although the electric-field-induced variation of DM interaction was recently reported for SrRuO$_3$/SrIrO$_3$[15], the role of the electric-field for DM interaction variation has yet to be clarified. Because of the short screening length of the electric-field in itinerant ferromagnet SrRuO$_3$, we expect that an electric-field applied to the ultrathin SrRuO$_3$/SrTiO$_3$ heterostructures may be utilized to efficiently manipulate the DM interaction due to the presence of additional Rashba effect[38,39], resulting in more interesting physical behaviors. In the investigation described below, the THE and its electrical control are observed for ultrathin SrRuO$_3$ single-layered films grown on SrTiO$_3$(001) substrates. In these samples, a peculiar pattern of oxygen octahedral tilting is observed by high-resolution lattice structure analysis, which adds to the inversion asymmetry in the SrRuO$_3$ layers right next to the SrTiO$_3$. The ensuing enhancement of the DM interaction is shown to stabilize the Néel-type magnetic skyrmion spin configuration, leading to THE-type transport. The present work opens a unique and promising window for the design of artificial materials with electrically tunable topological spin textures and the associated quantum transport properties.

**Results**

**Observation of THE in SrRuO$_3$ single-layered films.** High-quality SrRuO$_3$ epitaxial films with various thicknesses (4~50-unit cells, u.c.) were grown on insulating SrTiO$_3$(001) substrates using pulsed laser deposition (PLD) (see Methods,

Supplementary Fig. S1 and Note 1). Films were then patterned into Hall bar geometry devices for transport measurements with an applied current of 5 μA. The longitudinal resistivity ($\rho_{xx}$) versus temperature ($T$) curves of the SrRuO$_3$ films were measured from room temperature to 5 K. As the film thickness decreases, $\rho_{xx}$ continuously increases and three different regimes can be identified. For the thickness of 15~50 u.c., the films show a metallic behavior over the entire temperature range of 5~300 K. However, the $\rho_{xx}$ of films undergoes a metal-insulator-transition when the thickness $t$ is less than 10 u.c. Samples with a thickness of 4 u.c. or less display an insulating behavior in the entire temperature regime[40–42] (Supplementary Fig. S2 and Note 2). We show in Figs. 1a–d the magnetic field ($B$) dependent transverse Hall resistivity ($\rho_{yx}$) of the ultrathin SrRuO$_3$ films (10, 9, 8, and 5 u.c.). Note that the ordinary Hall contribution determined by the linear extrapolation in the high magnetic field region has been subtracted. For the 10 u.c.-thick SrRuO$_3$ sample (Fig. 1a), the $\rho_{yx}$-$B$ curve exhibits the typical anomalous Hall effect (AHE) signature in the temperature range of 10~100 K, indicating the perpendicular magnetic anisotropy of the SrRuO$_3$ film[3]. The Curie temperature ($T_C$) of 10 u.c. SrRuO$_3$ is ~138 K (Supplementary Figs. 2 and 3, Notes 2 and 3). A comparable AHE signal is also observed for the 9 u.c.-thick film (Fig. 1b), in which the ferromagnetic ordering at fixed temperature is a bit weaker than that of 10 u.c. sample because of the lower $T_C$ (Supplementary Fig. S2).

The situation changes dramatically when the film thickness is reduced to 8 u.c. (Fig. 1c). In stark contrast to conventional AHE of ferromagnetic conductors, the Hall traces exhibit a clear anomaly characterized by the appearance of a hump or dip structure superimposing on the usual AHE loop while scanning the field between −5 and 5 T. Similar features are also observed in the 5 u.c.-thick sample (Fig. 1d) when the temperature is below 70 K. This sample has an even lower $T_C$ (~95 K) (Supplementary Fig. S2). The hump/dip structure is the typical feature of the THE[14,15,21,26,27], which has a fundamentally different origin from AHE. It is proposed that a real space geometric phase acts on an electron as an emergent effective electromagnetic field through the interaction with the topological skyrmion spin textures[17]. Consequently, the moving electrons are scattered by the topological skyrmion spin textures in a direction opposite to the scenario of AHE, thus generating a topological Hall voltage[17,21,22]. THE has been recently observed in interfacial magnetic skyrmion systems as exemplified by $Cr_x(Bi_{1-y}Sb_y)_{2-x}Te_3/(Bi_{1-y}Sb_y)_2Te_3$ interface[21], $SrRuO_3/SrIrO_3$ interface[14,15], $CrTe/SrTiO_3$ interface[26], Mn-doped $Bi_2Te_3/SrTiO_3$ system[27], and exchange-biasing $Mn_4N/Mn_2N_y$[43] and $(Bi,Sb)_2Te_3/MnTe$[44] systems. The present work provides the first clear evidence for the THE in single-layered $SrRuO_3$ films grown on $SrTiO_3$(001) substrates.

Phenomenologically, the total Hall resistivity can be thought of as having three components[14,15]: the ordinary Hall effect, the AHE and the THE, expressed as

$$\rho_{yx} = R_0 B + R_S M + \rho_{THE}, \qquad (1)$$

where $R_0$ is the ordinary Hall coefficient, $R_S$ is the anomalous Hall coefficient, $\rho_{THE}$ is the topological Hall resistivity arising from topological skyrmion spin textures, and $B$ is the magnetic field perpendicular to the film plane. Compared with other three samples (Figs. 1a–c), the magnetization $M$ of 5 u.c.-thick film is lower but still sizable (Supplementary Fig. S4 and Note 4), and the sign of its AHE curves becomes positive (Fig. 1d), indicating the singularity of the band structure of $SrRuO_3$. This is possibly due to reduced $M$[34,42,45].

To has a close inspection of THE, we extract the THE contribution from the AHE curve at 10 K for the typical 5 u.c. $SrRuO_3$ film. As shown in Fig. 1e, the difference between the ascending (red) and descending (blue) branches in the saturated regime is shaded and marked as $\rho_{THE}$. Concerning the ascending branch, the $\rho_{yx}$ curve deviates from the normal AHE behavior at the critical field $B_{cr}$ of ~0.86 T and develops into a broad hump with a peak at a specific field $B_P$ of ~1.57 T. Above a characteristic field $B_T$ of ~2.65 T, the Hall curve merges back to the usual AHE hysteresis loop. This feature suggests that topological nontrivial spin structure with scalar spin chirality appear within 0.86~2.65 T. Based on the extracted THE from all the Hall traces at different temperatures, we build a skyrmion phase diagram for 5 u.c. SrRuO$_3$ sample, as represented by contour mapping of derived $\rho_{THE}$ value in the temperature $T$ and magnetic field $B$ planes, as displayed in Fig. 1f. The skyrmion phase indicated by the emergence of THE extends across a wide region in the $T$–$B$ plane. Within such a broad temperature and magnetic field ranges, the sign of $\rho_{THE}$ is always positive and varies systematically and reaches a maximum value of $\rho_{THE} \approx 0.30$ μΩ cm at 10 K. In a skyrmion system, each skyrmion traps an effective magnetic flux quantum $\phi_0 = h/e$, where $h$ is the Planck constant, $e$ is the elementary charge. The topological Hall resistivity induced by the skyrmion[14,15] can be written as

$$\rho_{THE} \approx PR_0 B_{eff}^z = PR_0 n_{sk} \phi_0, \qquad (2)$$

where $B_{eff}^z = n_{sk} \phi_0$ is the fictitious effective field, $P$ denotes the spin polarization of the conduction electron in SrRuO$_3$ and $n_{sk}$ is skyrmion density. Adopting the value of $R_0 = -0.08$ μΩ cm T$^{-1}$ measured at $T = 10$ K and the $P$ in the range from −9.5% to −50%[46,47], the skyrmion density $n_{sk}$ estimated from Eq. (2) varies from ~1.8 × 10$^{15}$ to ~9.5 × 10$^{15}$ m$^{-2}$. Consequently, the estimated size of a single skyrmion is $n_{sk}^{-1/2}$ ~ 10~23 nm. These values are quite reasonable compared with the typical diameter of DM interaction-induced skyrmion, which ranges from 5 ~ 100 nm[17,27].

**Electric-field control of THE.** An interesting issue is the manipulation of the THE.

We further investigated the electric-field control of the THE. The inset of Fig. 2a presents the schematic of the bottom gate dependence of measurements at 5 K. Because of the huge dielectric constant ($\varepsilon\sim23,000$)[48] of SrTiO$_3$ at low temperatures, an electrode of dried silver conductive paint on the bottom side of the SrTiO$_3$ substrate can form a field-effect transistor, which can be used to tune the chemical potential of the SrRuO$_3$ films. Figure 2a shows $V_g$ dependence of longitudinal resistivity $\rho_{xx}$. The variation of $\rho_{xx}$ ($\Delta\rho_{xx}/\rho_{xx}= ([\rho_{xx}(V_g) - \rho_{xx}(0)]/\rho_{xx}(0))$ is –3.2% (electron accumulation) under $V_g$ = 200 V and 3.5% (electron depletion) under $V_g$ = –170 V. It implies small or even negligible changes in carrier density (Fig. 2c), possibly due to the large carrier density of the intrinsic SrRuO$_3$ film[15,49].

In contrast to $\rho_{xx}$, the $V_g$ has a much stronger effect on AHE and THE. Figure 2b shows $\rho_H$-$B$ curves under the typical gate biases of −170, −100, 0, 100, and 200 V, where the ordinary Hall effect was not subtracted. Although the overall change of the slope (the ordinary Hall effect) at different $V_g$ is not so pronounced (coincide with the results in Fig. 2a), the AHE and THE parts are modulated within a large scale, as highlighted in Fig. 2d. The variation of AHE ($\Delta\rho_{yx}^0/\rho_{yx}^0 = [\rho_{yx}^0(V_g) - \rho_{yx}^0(0)]/\rho_{yx}^0(0)$) is 100% for $V_g$ = −170 V and 34.5% for $V_g$ = 200 V, which is at least one order of magnitude larger than the change of $\rho_{xx}$. Here $\rho_{yx}^0$ represents the zero magnetic field Hall resistivity.

Meanwhile, the THE (indicated by shaded area) exists under all electric biases but is most pronounced under $V_g$ = −170 V and greatly suppressed under $V_g$ = 200 V. The critical fields and the topological Hall resistivity are summarized in the $V_g$-$B$ plane (Fig. 2e). The broad shaded area between the $B_{cr}$- and $B_T$-$V_g$ lines represents the skyrmion phase. The peak values of $\rho_{THE}$ are 0.52, 0.43, and 0.28 μΩ cm under $V_g$ = −170, 0, and 200 V, respectively. The corresponding variation in peak of $\rho_{THE}$ ($\Delta\rho_{THE}/\rho_{THE} = [\rho_{THE}(V_g) - \rho_{THE}(0)]/\rho_{THE}(0)$) is 21% under $V_g = -170$ V and −35% under $V_g = 200$ V. As a consequence, the size of individual topological skyrmion spin texture, estimated from Eq. (2), is ~8 nm ($V_g = -170$ V) and ~10.5 nm ($V_g = 200$ V) adopting the spin polarization of $P = -9.5\%$[46]. The size of the skyrmion scatters in the range of 10~30 nm when $P$ varies from −20% to −50%[14,47] (Fig. 2h). Therefore, the THE in the single-layered SrRuO$_3$ is robust, occurring in a wide range of temperatures, magnetic fields, and gate voltages.

A comparison of the $V_g$ control of carrier density and AHE/THE suggests that the observed large modulation of AHE and THE cannot be simply explained by the $V_g$-induced carrier density change. Possibly, the SOC may suffer from the tuning of electric gate, resulting in a variation in DM interaction. We note that the coercive field ($H_C$) of the AHE curves monotonically changes from 1.19 T ($V_g = 200$ V) to 1.07 T ($V_g = 0$), and then to 0.94 T ($V_g = -170$ V) at 5 K (Supplementary Fig. S5 and Note 5). Meanwhile, AHE also exhibits a dramatic variation under gate biases. Both phenomena are closely related to SOC thus support the modulation of SOC[15].

**Atomic-scale characterization of the interfacial oxygen octahedral tilting.** The transport experiments have revealed the robust THE and electric modulation of THE in ultrathin SrRuO$_3$ single-layered films. An atomic scale analysis of the interfacial structures is required to reveal the physical origin of THE. The atomic scale elemental mapping and the corresponding high-angle annular dark field scanning transmission electron microscopy (HAADF-STEM) and annular bright-field (ABF)-STEM images of the 8 u.c. SrRuO$_3$ films on a SrTiO$_3$ substrate along the [010]$_O$ zone axis of SrRuO$_3$ with an orthorhombic structure are shown in Figs. 3a–c, respectively. Atomic columns of SrO (green), RuO (red) and TiO (blue) are clearly distinguished from each other based on column intensity differences, clearly revealing a coherent film/substrate interface without any misfit dislocations. The composite color map of Ru atoms in red and Ti atoms in blue shows the atomically sharp interface without any elemental intermixing, which is further confirmed by the absence of peak overlapping between Ti $L_{2,3}$ edges and Ru $M_{2,3}$ edges across the SrRuO$_3$/SrTiO$_3$ interface, as shown by the atomic scale electron energy loss spectroscopy (EELS) (Supplementary Fig. S6). The O $K$ edges and Ti $L_{2,3}$ edges of eight atomic layers at the interface are almost identical, indicating the same valence states of Ti and the absence of oxygen vacancies. Surprisingly, the oxygen atom positions arranged in a zigzag pattern along transverse direction in a few layers of orthorhombic-like SrRuO$_3$ (O-SRO) at interface are shown in the ABF-STEM image in Fig. 3c, demonstrating the interfacial oxygen octahedral tilting of SrRuO$_3$. As shown in the insets of Fig. 3d, the magnified ABF-STEM image of the first atomic layer of SrRuO$_3$ at the very interface is consistent with the simulated ABF-STEM image of O-SRO along the $[010]_O$ zone axis under the same experimental conditions. The oxygen octahedral tilting of the SrRuO$_3$ layers close to the interface is smaller than that of bulk orthorhombic SrRuO$_3$[50,51], so the layers of SrRuO$_3$ with the oxygen octahedral tilting close to the interface are described as orthorhombic-like structure.

Figure 3d shows the variations of oxygen octahedral tilting angles at each atomic layer across the SrRuO$_3$/SrTiO$_3$ interface by statistical analysis of each RuO and TiO atomic columns. For instance, the average tilting angle of the first atomic layer is about 3.16 ° with a standard deviation of 0.15 °, by analyzing the tilting angles of all the atomic columns (Supplementary Fig. S7). The oxygen octahedral tilting angle of each atomic plane in SrRuO$_3$ along the $[010]_O$ zone axis dramatically decreases with the increased distance from interface, and there is no oxygen octahedral tilting for the tetragonal SrRuO$_3$ (T-SRO) above the 5th atomic layer. However, the spatial changes of oxygen octahedral tilting here close to the interface region could directly break the inversion symmetry, which contributes to the DM interaction. An epitaxial orientation relationship of SrTiO$_3$[110](001)//O-SRO[010](001)//T-SRO[010](001) is further confirmed by investigating the samples along [110]$_O$, [1$\bar{1}$0]$_O$ and [100]$_O$ zone axes (Supplementary Figs. S8 and S9). No oxygen octahedral rotation and tilting of T-SRO and O-SRO is observed along these three zone axes, which is consistent with the corresponding project of the perfect crystal (Supplementary Figs. S8 and S9). Note that the ultrathin SrRuO$_3$ undergoes an orthorhombic-to-tetragonal phase transition. Furthermore, the tetragonality becomes robust with decreasing SrRuO$_3$ film thickness. The intrinsic tetragonal phase is also supported by the X-ray reciprocal space mappings (RSMs) of 16 u.c. and 8 u.c. SrRuO$_3$ films (Supplementary Figs. S10 and S11 and Note 6). Our finding coincides with the previously reported results that the tetragonal SrRuO$_3$ phase can be epi-stabilized at room temperature for $t \leq 17$ u.c.[52].

In general, the octahedral distortions in ultrathin hybrid structure are expected to accommodate the symmetry-mismatch between the film and substrate with the coherent interface. The corner connectivity of octahedron across the interface drives the film to adopt the tilting symmetry of substrates, which tends to suppress the octahedral tilts (cost energy). Meanwhile, mechanical rigidity of film octahedron tends to develop its own tilting symmetry away from the interface (gain energy)[50,53,54]. Consequently, the final rotation and tilting structures of ultrathin hybrid structure are determined by the equilibrium of the energy states between these two competing effects. For the ultrathin SrRuO$_3$ films here, due to the intrinsic robust tetragonal inclination of SrRuO$_3$ itself and the cubic nature of the SrTiO$_3$ substrate, the neighboring interfacial RuO$_6$ octahedra cooperatively adjust their orientation with $c$-axis out-of-plane to follow the growth mode of bottom TiO$_6$ octahedra and connect upper tetragonal RuO$_6$ octahedra ($c$-axis out-of-plane) in such a manner to develop its own $a^-a^-c^+$ tilting symmetry. As an ultimate result there is a octahedral tilting along [010]$_O$ zone axis but the tilting angle is suppressed to ~3.16°. Afterwards, the orthorhombic RuO$_6$ octahedra with $c$-axis out-of-plane deformation rapidly becomes non-tilting tetragonal symmetry above several unit cell layers. It is then deduced that the pseudo-tetragonal 8 u.c. film is more energetically stable than forming a bulk-like tilting structure. Similar results were observed in ultrathin (La,Sr)MnO$_3$/(LaAlO$_3$)$_{0.3}$(LaSrTaO$_6$)$_{0.7}$ and LaCoO$_3$/SrTiO$_3$[53,54].

In addition to the strong SOC readily available in 4$d$ oxide SrRuO$_3$, the other prerequisite for a finite DM interaction is the broken inversion symmetry. This naturally occurs for the asymmetric oxygen octahedron condition between the reconstructed interfacial tilting RuO$_6$ octahedrons and non-tilting RuO$_6$ octahedrons away from the interface induced by the local orthorhombic-to-tetragonal structure phase transition. Thus, the interfacial oxygen octahedral tilting here is highly expected to be responsible for the emergence of THE.

**Control of THE by interfacial engineering.** We then turn towards the control experiment of manipulating THE by interfacial engineering. To further confirm the crucial role of the interfacial octahedral tilting on the observed THE, a 2~4 u.c. insulating and nonmagnetic BaTiO$_3$ buffer layer is intentionally inserted between SrRuO$_3$ and SrTiO$_3$ substrates to form SrTiO$_3$/BaTiO$_3$ ($N$ u.c.)/SrRuO$_3$ (8 u.c.) heterostructures ($N$ = 2, 3, and 4 u.c.) (see Methods, Supplementary Fig. S1 and Note 1). For the $N$ = 2 heterostructure, remarkably, a small, but apparent THE signal (indicated by black arrows) remains at different temperatures from 5 to 70 K (Fig. 4a).

The scenario differs dramatically for the heterostructures with $N = 3$ and $N = 4$. Concomitant $\rho_{yx}$-$B$ curves at various temperatures are shown respectively in Figs. 4b and c, where the THE completely disappears. These results suggest that the interfacial $RuO_6$ octahedral tilting is suppressed and/or blocked by the $BaTiO_3$ insertion, resulting in the absence of inversion symmetry breaking. We also observed that the sign of $\rho_{yx}$ is inverted from negative to positive when the $BaTiO_3$ is 3 and 4 u.c.-thick. The sign reversal of AHE is likely due to the singularity in the band structure of $SrRuO_3$ through the Ru-O-Ru bond angle changes[34]. Previous studies reported that the Ru-O-Ru bond angle changes from 168° at the $SrRuO_3$/$GdScO_3$ interface to 180° at the $BaTiO_3$/$SrRuO_3$ interface as the $BaTiO_3$ inserting or capping layer was presented[55]. This would lead to the variation of Berry curvature and resultant band crossings in $SrRuO_3$.

In order to further demonstrate the absence of interfacial octahedral tilting due to the introduction of $BaTiO_3$, the HAADF-STEM and ABF-STEM images of the epitaxial $SrTiO_3$/$BaTiO_3$ (4 u.c.)/$SrRuO_3$ (8 u.c.) thin films were acquired as shown in Figs. 4d and e, respectively. In comparison with the first $SrRuO_3$ layer at the $SrRuO_3$/$SrTiO_3$ interface, there is no significant octahedral tilting of the first $SrRuO_3$ layer at the $BaTiO_3$/$SrRuO_3$ interface as shown in the Supplementary Figs. 7c and d. Figure 4f shows the variations of oxygen octahedral tilting angles at each atomic layer across the $SrTiO_3$/$BaTiO_3$/$SrRuO_3$ interface by statistical analysis, clearly revealing no octahedral tilting in the entire thin films.

**Calculated DM interaction and skyrmion configuration**. THE is thought to arise from non-trivial spin textures, especially the topological skyrmions, induced by DM interaction. To clarify the microscopic origin of the experimentally observed THE and its variation under the applied perpendicular electric field in the 5 u.c.-thick $SrRuO_3$ film grown on $SrTiO_3$ substrate, the DM interaction is quantitatively determined based on a tight binding Hamiltonian, and the skyrmion configuration is subsequently obtained. The DM interaction originates from the SOC and spatial inversion symmetry breaking[17,56]. In our experiment, the inversion symmetry of $SrRuO_3$ film is removed by three factors, namely, the proximity to $SrTiO_3$ substrate, the interfacial oxygen octahedral tilting induced by local orthorhombic-to-tetragonal structural phase transition of $SrRuO_3$ film and, lastly, the applied perpendicular electric field. These three factors are thus expected to contribute to DM interaction, and are described by a tight binding Hamiltonian $H = H^{hop} + H^{soc} + H^R$ incorporating Ru $t_{2g}$ orbitals. $H^{hop}$, $H^{soc}$ and $H^R$ correspond to the hopping term, the atomic-like spin-orbit interaction and the Rashba interaction respectively. The Rashba interaction arising from the application of the perpendicular electric field results in the change of electrons orbital characters in hopping processes[57–59]. The strength of the Rashba interaction is represented by $\lambda$ (see Methods and Supplementary Note 7).

In the continuum limit, the energy of the 2D system is a functional of the magnetization field through both exchange interaction $J$ and the DM interaction $D$,

$$E = \frac{1}{S}\int dr J(\nabla \boldsymbol{m})^2 + \sum_\mu \boldsymbol{D}_\mu \cdot (\partial_\mu \boldsymbol{m} \times \boldsymbol{m}), \quad (3)$$

where $\boldsymbol{m}$ is the magnetization direction at $\boldsymbol{r}$, $S$ is the integral area, and vector $\boldsymbol{D}_\mu$ parametrizes effective DM interaction with $\mu = x, y, z$. The strength of $J = 0.71 \text{ meV}/a_0$ is obtained according to the measured $M$-$T$ relation, where $a_0$ is the nearest Ru-Ru distance (see Supplementary Fig. S12 and Note 7). The coefficients of DM interaction $\boldsymbol{D}_\mu$ are calculated from the spin susceptibility based on the tight-binding model[60]. The structure of an isolated skyrmion can be obtained by minimizing the energy in Eq. (3)[21,27].

We first examine the effects of SrTiO$_3$ substrate and the interfacial oxygen octahedral tilting induced by structural phase transition. The SrTiO$_3$ substrate induces on-site energy difference about 31 meV between the $d_{zx}$ ($d_{yz}$) orbitals of 1$^{st}$ layer Ru atoms and the 2$^{nd}$ layer Ru atoms close to the substrate, based on our density-functional theoretic calculations. The calculated nonzero components of DM interaction for SrRuO$_3$ film with tetragonal structure are $D_{xy} = -D_{yx} = 0.052 \text{ meV}/a_0^2$, in the absence of oxygen octahedral tilting and the external perpendicular electric field. The diameter of isolated skyrmions, upon energy minization, is around 84 nm, which is much larger than the estimated value (10~30 nm) from the experimentally measured THE. Therefore, the observed THE cannot be attributed entirely to the SrRuO$_3$/SrTiO$_3$ interface effect, and seem to be enhanced by additional DM interaction. Indeed, such enhancement comes from the interfacial oxygen octahedral tilting induced by local orthorhombic-to-tetragonal structural phase transition of SrRuO$_3$ (Fig. 3).

The local orthorhombic-to-tetragonal phase transition is presented in the SrRuO$_3$ film as shown in Fig. 5a, where the oxygen octahedral tilting of the 1$^{st}$ and 2$^{nd}$ SrRuO$_3$ layer close to SrTiO$_3$ substrate highlighted by a blue rectangular box. According to the STEM measurement (Supplementary Fig. S7), the tilting angles are about 4 ° for the 1$^{st}$ layer SrRuO$_3$ and 2 ° for the 2$^{nd}$ layer SrRuO$_3$. When these seemingly minute structural modulations are considered in the tight-binding model, the computed DM interaction is shown in Figs. 5b and c ($\lambda = 0$ case). The obtained skyrmion configuration is Néel-type[21,27] with a diameter of 22 nm (Fig. 5f), in good agreement the experimental estimate. Therefore, the interfacial oxygen octahedral tilting induced by local structural phase transition plays a critical role in the formation of skyrmion and the observed THE. Starting with the optimal continuum magnetization field, we further relax the skyrmion structure over a square grid with fixed boundary to minimize the energy density using conjugate-gradient method. The computed configuration of Néel-type skyrmion (Fig. 5g) remains essentially unchanged.

Finally, the electric-field control of THE is numerically simulated in the presence of interfacial oxygen octahedral tilting. The strength of electric field is approximately proportional to the Rashba interaction strength $\lambda$. Figs. 5b and c display the numerical results of DM interaction as functions of $\lambda$, showing that the components of DM interaction are remarkably sensitive to the variation $\lambda$. The determinant $|D| = D_{xx}D_{yy} - D_{xy}D_{yx}$ is always positive, indicating that the system favors to form skyrmions rather than anti-skyrmions[61], over the range of electric field considered. The energy minima of skyrmion and the skyrmion size as a function of $\lambda$ are presented in Figs. 5d and e, respectively. As $\lambda$ changes from negative to positive, the energy density becomes higher and the size of skyrmion turns out to be larger. This is in good agreement with the experimental measurements of THE, which gives larger size of skyrmion at positive gate voltages. The perpendicular electric field at the ultrathin $SrRuO_3/SrTiO_3$ interface further breaks the inversion symmetry causing Rashba interaction, which is observed as electrical control of Rashba-type band splitting in an inverted $In_{0.53}Ga_{0.47}As/In_{0.52}Al_{0.48}As$ semiconductor heterostructure[38] and $LaAlO_3/SrTiO_3$ oxide interface[62]. This factor leads to a sizeable interface DM interaction modulation and ultimately leads to the observed electric-field induced modulation of THE.

**Discussion**

To summarize, we demonstrate the existence of THE in ultrathin 4$d$ $SrRuO_3$ ($\leq 8$

u.c.) films grown on $SrTiO_3$(001) substrates. The THE can be greatly modulated by an electric-field. The physical origin of the THE in $SrRuO_3$ can be understood as follows: A combination of broken inversion symmetry induced by $RuO_6$ octahedral tilting at the $SrRuO_3/SrTiO_3$ interface region and strong spin-orbit interaction of $SrRuO_3$ generate chiral DM interaction. It is the DM interaction that stabilizes Néel-type magnetic skyrmion spin texture in $SrRuO_3$ single-layered films. It is known that a spin texture with multi-**q** spiral can also lead to THE[63]. In our calculations, we have also assayed possible spiral spin textures with a pair of **q** vectors. Upon energy minimization, these spiral structures lead to a square lattice of skyrmions with identical chirality and similar size as in the single-skyrmion model discussed above. Therefore, our discussion based on the formation energy of single skyrmion provides an adequate model for the system.

The main contributions of the present work are the following: (i) The $SrRuO_3$ single-layered films can generate the THE; (ii) The realization of electric-field control of DM interaction paves an avenue for electrically tuning the chiral spin structures, which has been proposed with potential applications for high-density and low power consumption memories; (iii) Oxygen octahedral tilting not only serves as an intrinsic mechanism for the THE in oxides system, but also plays a role to produce and/or modulate DM interaction in oxide films. Such a bridge will connect the oxide community and spintronics, as well as topological materials community to explore more interesting phenomenon based on the oxides. No observation of skyrmions by real-space spin-sensitive imaging techniques (e.g. Lorenz transmission electron microscopy (LTEM)) is likely due to the too small skyrmion size in the $SrRuO_3$ films. Our observation of the THE and its electrical modulation feature would facilitate further studies of the topological skyrmion spin textures in more oxide materials.

*Note added.*—After we submitted our manuscript for consideration on September 6th, 2018, we found an interesting and independent work reporting the emergence of skyrmions and THE in ultrathin SrRuO$_3$ films on SrTiO$_3$ on October 4th, 2018.[64] The authors assumed that the octahedral tilting angle is zero for ultrathin SrRuO$_3$ films on SrTiO$_3$, so the rumpling of Ru-O plane was proposed to be the source of DM interaction. Our work demonstrated the RuO$_6$ octahedral tilting-induced inversion symmetry breaking and DM interaction and resultant THE in ultrathin SrRuO$_3$ films on SrTiO$_3$ substrates. The interfacial RuO$_6$ octahedral tilting induced by local orthorhombic-to-tetragonal structural phase transition is directly observed by STEM.

**Methods**

**Sample preparation**

High-quality SrRuO$_3$ films, BaTiO$_3$/SrRuO$_3$ heterostructures were grown by pulsed laser deposition (PLD) with a KrF excimer laser from stoichiometric SrRuO$_3$ and BaTiO$_3$ targets in layer-by-layer mode on the (001)-oriented SrTiO$_3$ substrates (5 mm × 5 mm × 0.5 mm). The SrRuO$_3$ was grown at 700 ℃ with an oxygen background pressure of 100 mTorr and a repetition rate of 5 Hz, while BaTiO$_3$ was grown at 720 ℃ with an oxygen background pressure of 4 mTorr and a repetition rate of 2 Hz. The growth was *in situ* monitored by reflection high-energy electron diffraction (RHEED). This allows precise control of the film thickness at the unit cell scale and accurate characterization of the growth dynamics. After growth process, the samples were slowly cooled down to room temperature in 300 Torr of oxygen pressure at a rate of ~5 ℃/min to improve the oxidation level.

The atomic arrangements of SrRuO$_3$ have two typical structures of tetragonal and orthorhombic phases. In the case of tetragonal SrRuO$_3$, it possesses $a^0a^0c^-$ rotation, which leads to no oxygen octahedral rotation along $[010]_T$ direction and no oxygen octahedral tilting along $[110]_T$ direction (see Supplementary Fig. S9). Conversely, the orthorhombic SrRuO$_3$ possess $a^-a^-c^+$ rotation, which results in oxygen octahedral rotation along $[001]_O$ direction and octahedral tilting along $[010]_O$ direction, but oxygen atoms are arranged in a zigzag way in the depth direction along $[100]_O$, $[110]_O$ and $[1\bar{1}0]_O$ zone axes (see Supplementary Fig. S9). Hence, along these three zone axes, the oxygen octahedral rotation and tilting cannot be visualized due to the limitation of spatial resolution in STEM. For convenience, all of crystal index have been indexed by the indices of orthogonal and tetragonal phase. The Glazer symbol[65] used above was referred to the rotation along $[1\bar{1}0]_{O/T}$, $[110]_{O/T}$, and $[001]_{O/T}$ respectively.

**Crystal structural characterization**

The X-ray reciprocal space mappings of the films were measured by using a Bruker X-ray diffractometer equipped with thin film accessories (D8 Discover, Cu K$_\alpha$ radiation) at the Beijing National Laboratory for Condensed Matter Physics, Institute of Physics, Chinese Academy of Sciences.

**Magnetization and transport measurements**

A superconducting quantum interference device (SQUID) magnetometer with a magnetic field applied perpendicular to the film plane was used to measure the magnetic properties of samples because the magnetic easy axis of the SrRuO$_3$ film is nearly perpendicular to the film plane when grown on SrTiO$_3$(001) substrate. A Hall-bar geometry device was used to carry out the electrical transport measurements in a 9 Tesla mini Cryogen-Free Measurement System for Precision Measurement of Physical Properties (CFMS-9, Cryogenic Limited, United Kingdom). The sample was patterned into a Hall-bar by photolithography and Ar ion etching process for Hall and longitudinal resistivity measurements. The effective channel is 400 μm long and 100 μm wide. All Ti (20 nm)/Au (80 nm) electrodes were deposited using electron-beam evaporation, resulting in low-resistant Ohmic contacts. Magnetic field dependence of Hall resistivity was obtained by subtracting the ordinary Hall contribution by linear extrapolation in the high magnetic field region. Antisymmetric process was performed for all the Hall data shown in this paper.

**Scanning transmission electron microscopy (STEM)**

The cross-section TEM lamellas of the $SrTiO_3/SrRuO_3$ (8 u.c.) and $SrTiO_3/BaTiO_3$ (4 u.c.)/$SrRuO_3$ (8 u.c.) epitaxial thin films along the desired zone axes were prepared by using a Zeiss dual-beam focused ion beam (FIB). The thin films were capped with a thin Pt film to prevent from being damaged during sample preparation of TEM cross-section specimen. The final thinning step was performed with 2 kV 1mA Ar ion milling to remove the amorphous layer. STEM analysis was performed on a double aberration-corrected microscope FEI Titan G3 electron microscope equipped with an aberration corrector for the probe-forming lens and a high-brightness gun, as well as a super-EDX 4-quadrant detector, operated at 300 kV acceleration voltage to acquire HAADF-, ABF-STEM images and conduct the EDX experiment. The setup for STEM convergence semi-angle used was 25 mrad, providing a probe size of ~0.6 Å. The collection semi-angle from 70–160 mrad and 11–29 mrad for HAADF and ABF imaging, respectively. Dual EELS mode is used to simultaneously acquire both the zero-loss and core-loss EELS spectra to ensure intrinsic chemical shift compensated the instrument drift by the correction of zero-loss. And atomic scale line scan of EELS spectra across the interface is acquired with a higher energy dispersion (0.1 eV/channel) to obtain Ti $L_{2,3}$- and O $K$-edge EELS.

**The tight binding model of SrRuO$_3$**

The $t_{2g}$ orbitals of Ru atoms are used to construct the tight binding model. The tight-binding Hamiltonian $H = H^{\text{hop}} + H^{\text{soc}} + H^R$ is composed of hopping part, SOC and Rashba interaction. Following the previous work[57,66], the nearest-neighbor in-plane and out-of-plane hopping parameters are $t_1 = 0.38\,\text{eV}$ and $t_2 = -0.2\,t_1$, respectively. The next nearest-neighbor hopping element is $f = 0.2\,t_1$. The atomic SOC strength $\lambda^{\text{soc}} = 0.4\,t_1$ and the spin splitting $m = -0.48\,t_1$.

To model the local rotations of the RuO$_6$ octahedrons, we rewrite the SOC term $H^{\text{soc}}$ from the local orientation-dependent basis to the global basis. According to the STEM measurements, the topmost 3 layers of SrRuO$_3$ possess a tetragonal structure with $a^0 a^0 c^-$ rotations (Fig. 3), while the bottom two layers exhibit a local tilting and develop an orthorhombic structure with $a^- a^- c^+$ rotations. The tilting angle is regarded as 4° for the 1$^{\text{st}}$ layer close to the interface and 2° for the 2$^{\text{nd}}$ layer close to the interface. The in-plane rotation angle is treated as 11.4° (Ref. 67).

To estimate the interface effect between SrRuO$_3$/SrTiO$_3$. we performed the DFT calculations for SrRuO$_3$ (5 u.c.)/SrTiO$_3$ (5 u.c.) supercell with the Vienna *ab initio* simulation package (VASP)[68]. The electron hopping matrix elements are constructed using Wannier90 code[69], incorporating Ru $d$ orbitals and Ti $t_{2g}$ orbitals. The SrTiO$_3$ substrate results in on-site energy difference about 31 meV between the $d_{zx}$ ($d_{yz}$) orbitals of the two bottom Ru layers, where on-site energy of electrons on the 1$^{\text{st}}$ layer (close to the interface) is higher.

The external perpendicular electric field induces an extra Rashba term[58,59]

$$H^R = \lambda \begin{pmatrix} 0 & 0 & -2i\sin k_x \\ 0 & 0 & -2i\sin k_y \\ 2i\sin k_x & 2i\sin k_y & 0 \end{pmatrix},$$

in the basis ($d_{yz}$, $d_{zx}$, $d_{xy}$). Here $\lambda$ is the strength of the Rashba interaction. Given that the electric field strength drops quickly from the interface to the itinerant ferromagnet SrRuO$_3$ layers, the Rashba term is only added to the 1$^{st}$ SrRuO$_3$ layer close to the interface.

**Skyrmion configuration**

The common form of skyrmion configuration[21,27] we used is

$$\boldsymbol{m} = (\sin(1-r/R)\cos(\theta+\phi), \sin(1-r/R)\sin(\theta+\phi), \cos(1-r/R)),$$

where $\boldsymbol{r} = (r\cos\theta, r\sin\theta)$ and $R$ is the radius of skyrmion, and $\phi$ is the phase factor characterizing the type of skyrmion. The skyrmion configuration can be determined by minimizing the energy density with respect to the radius $R$ and the phase $\phi$. The competition between ferromagnetic, spiral and skyrmion states is discussed in Supplementary Fig. S12b and Note 7.

**Acknowledgements**

The authors thank Dr. Chang Liu for fruit discussions. C.S. acknowledges the support of Beijing Innovation Center for Future Chip (ICFC) and Young Chang Jiang Scholars Program. This work was supported by the National Key R&D Program of China (No. 2017YFA0206302), the National Natural Science Foundation of China (Grant Nos. 51671110, 51571128, 51761135131, 51822105, 11725415 and 11804118), and Strategic Priority Research Program of Chinese Academy of Sciences (Grant No. XDB28000000). F.W. and C.Z.C. thank the support from Alfred P. Sloan Research Fellowship and ARO Young Investigator Program Award (W911NF1810198). This work made use of the resources of the National Center for Electron Microscopy in Beijing for characterizations and the Tianhe-I Supercomputer System for a part of calculations.

**Author contributions**

C. S., J.F., X.Y.Z. and Z.D.Z. conceived the experiment and directed the overall project. Y.D.G. prepared the samples, measured the transport properties, processed the experimental data. Y.W.W. and J.F. carried out the theoretical calculations. K.X., X.Y.Z. and J.Z. performed the STEM, EDX and EELS characterization and analysis. H.R.Z. and J.R.S. contributed to the X-ray RSM measurements. F.W., F.L., M.S.S., C.Z.C., W.L. and F.P. provided advice on the experiments. Y.D.G, Y.W.W., K.X., X.Y.Z., J.F., and C.S. interpreted the data and wrote the manuscript. All of the authors contributed to the extensive discussion of the results and commented on the manuscript.

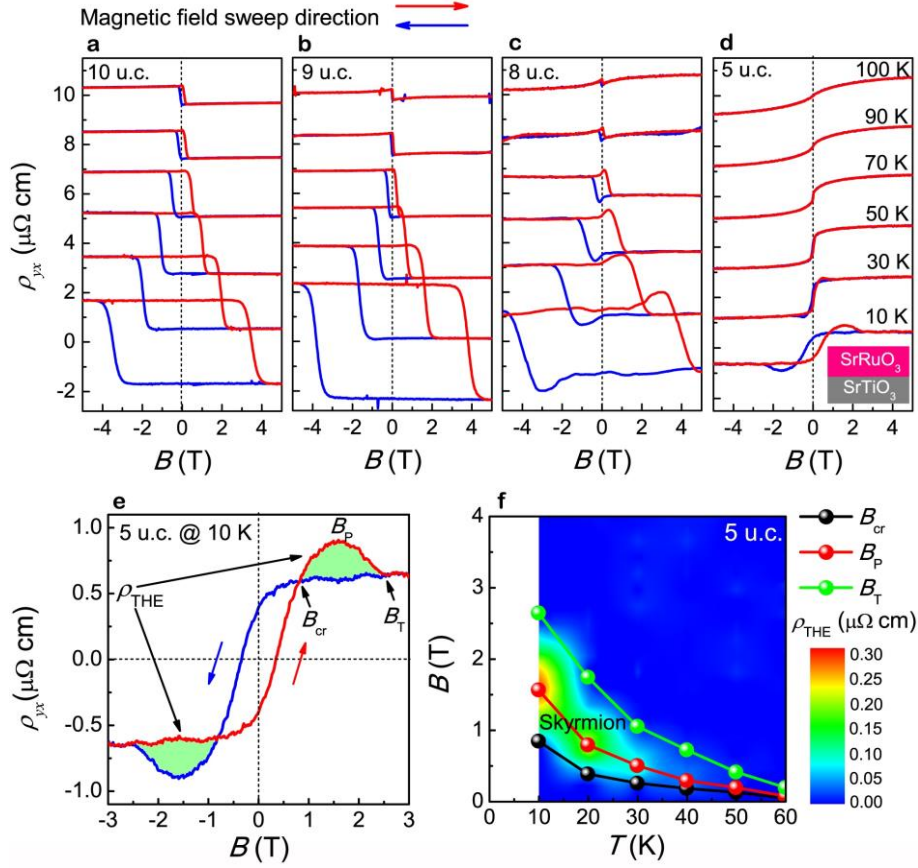

**Fig. 1 | THE in SrRuO₃ single-layered films.** $\rho_{yx}$-$B$ curves of single-layered SrRuO₃ films with various film thicknesses, $t$ = 10 u.c. (**a**), 9 u.c. (**b**), 8 u.c. (**c**), 5 u.c. (**d**) in the temperature range of 10~100 K. The red (blue) curve represents the ascending (descending) branch. The graphs have been shifted along the vertical axes for clarity. **e**, Typical $\rho_{yx}$-$B$ curve of 5 u.c. SrRuO₃ film at 10 K to highlight the THE. The difference between the ascending (red) and descending (blue) branch is shaded and marked as $\rho_{THE}$. **f,** The contour mapping of extracted topological Hall resistivity $\rho_{THE}$ of 5 u.c. SrRuO₃ film as a function of external magnetic field ($B$) and temperature ($T$). The hump area between the $B_{cr}$ (field for the appearance of THE) and $B_T$ (field for the disappearance of THE) lines represents the skyrmion regime. THE reaches its maximum value at the field of $B_P$.

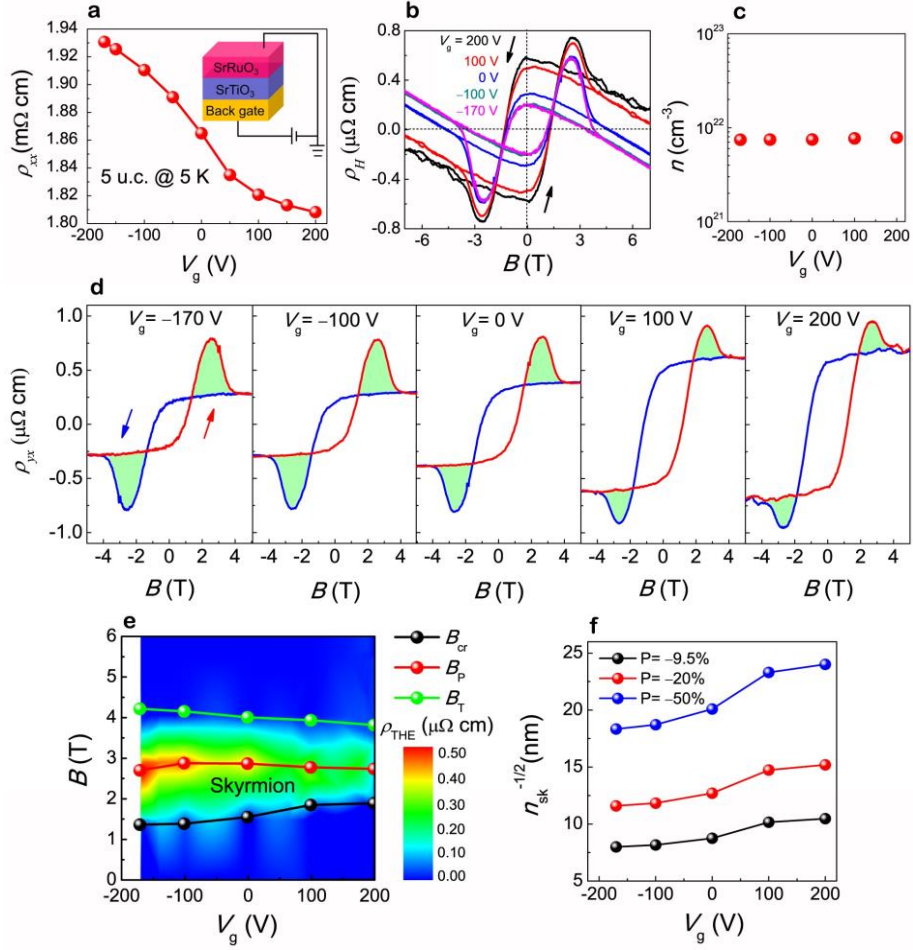

**Fig. 2 | Electric-field control of THE. a**, Gate voltage $V_g$ dependence of the longitudinal resistivity $\rho_{xx}$ of 5 u.c. SrRuO$_3$ films at 5 K. Inset: the schematic of bottom gate dependence of transport measurements using the SrTiO$_3$ as the dielectric layer. **b**, Magnetic field dependence of Hall resistivity $\rho_H$ (original data with the ordinary Hall effect) of 5 u.c. SrRuO$_3$ films at 5 K for $V_g = -170, -100, 0, 10,$ and 200 V. **c**, Carrier density $n$ plotted as a function of the gate voltage $V_g$. **d**, $V_g$ dependence of $\rho_{yx}$-$B$ curves after subtracting the ordinary Hall effect. **e**, Contour mapping of $\rho_{THE}$ in the $B$-$V_g$ plane at 5 K for 5 u.c.-thick SrRuO$_3$ films. The broad area between the $B_{cr}$ and $B_T$ lines represents the skyrmion regime in the whole $V_g$ ranges. THE reaches its maximum value at the field of $B_P$. **f**, $V_g$ dependence of the skyrmion size ($n_{sk}^{-1/2}$) based on three selective spin polarization values of SrRuO$_3$.

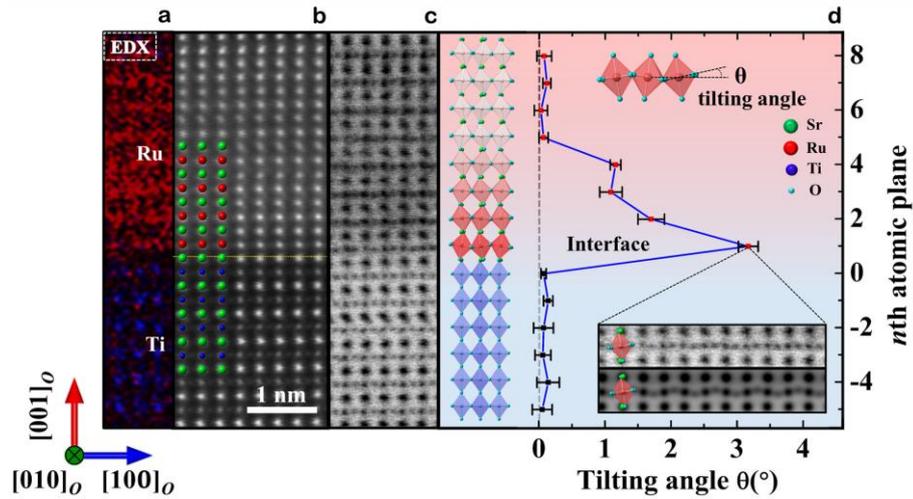

**Fig. 3 | Atomic-scale characterization of interfacial RuO$_6$ oxygen octahedral tilting across the SrRuO$_3$/SrTiO$_3$ (8 u.c.) interface.** The Ru and Ti elemental mapping (**a**) HAADF-STEM, (**b**) and ABF-STEM (**c**) images of 8 u.c. SrRuO$_3$ films on the SrTiO$_3$ substrate demonstrate atomically sharp interface without any element intermixing. The atomic positions of oxygen atoms in **c** clearly reveal the oxygen octahedral tilting along transverse direction at a few RuO$_6$ layers. **d**, Statistic analysis of depth-dependent oxygen octahedral tilting angles with monolayer resolution across the SrRuO$_3$/SrTiO$_3$ (8 u.c.) interface with the corresponding schematic models. The inset shows the magnified ABF-STEM images of the first SrRuO$_3$ layer at the interface and the simulated ABF-STEM image of orthorhombic-like SrRuO$_3$ under the same experimental conditions.

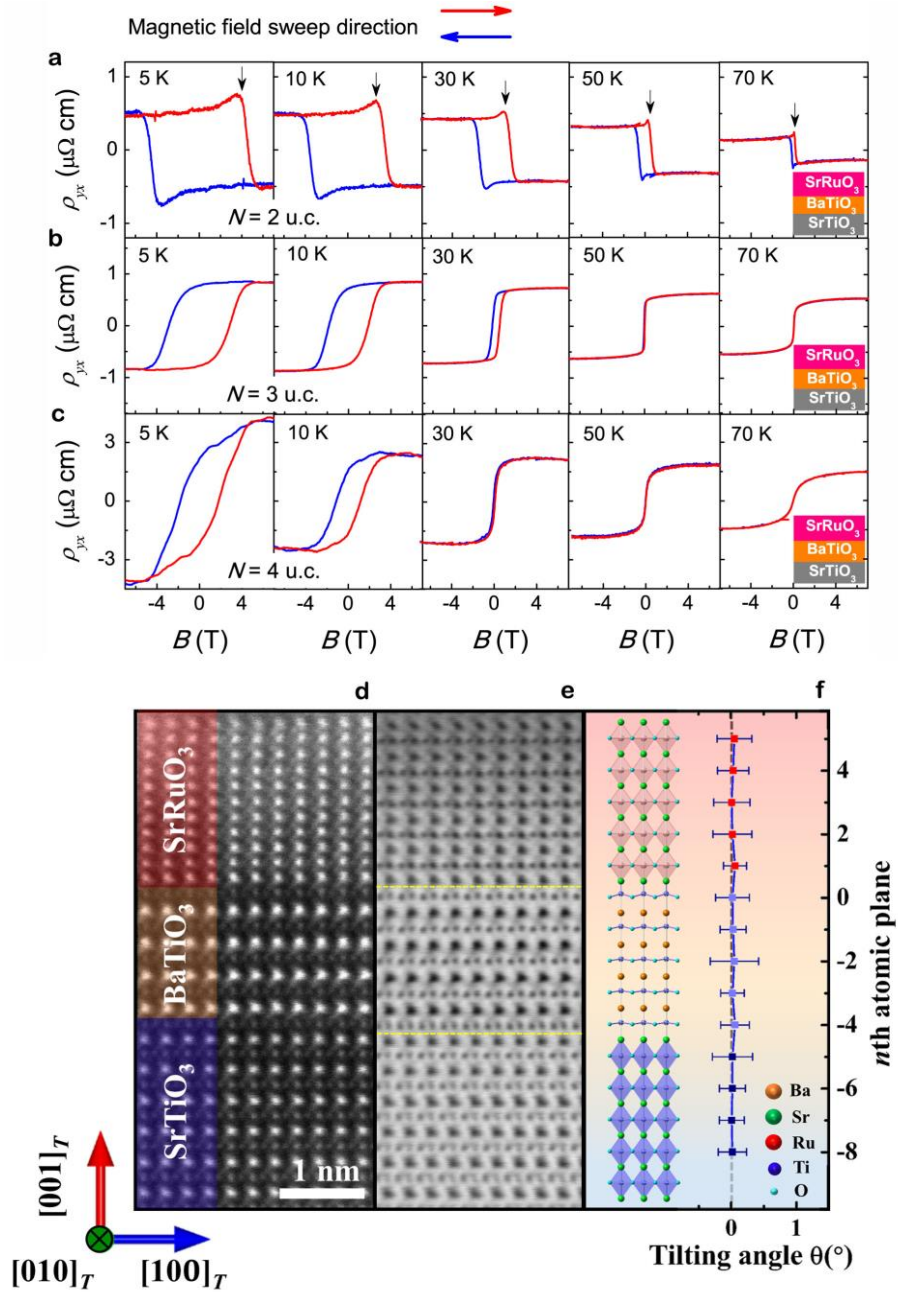

**Fig. 4 | Control of THE via interfacial engineering.** $\rho_{yx}$-$B$ curves of SrTiO$_3$/BaTiO$_3$ ($N$ u.c.)/SrRuO$_3$ (8 u.c.) [$N$ = 2 u.c. (**a**), 3 u.c. (**b**) and 4 u.c. (**c**)] ($N$: number of monolayers in BaTiO$_3$ layer) heterostructures at various temperatures. Blue (red) curve represents the process for decreasing (increasing) magnetic field. Ordinary Hall contribution is subtracted by linear extrapolation in the high magnetic field region. The HAADF-STEM (**d**) and ABF-STEM (**e**) images of the representative SrTiO$_3$/BaTiO$_3$ (4 u.c.)/SrRuO$_3$ (8 u.c.) heterostructures clearly show the atomically

sharp interface without any oxygen octahedral tilting. **f**, Statistical analysis of depth-dependent oxygen octahedral tilting angles with monolayer resolution shows no oxygen octahedral tilting across the entire thin films.

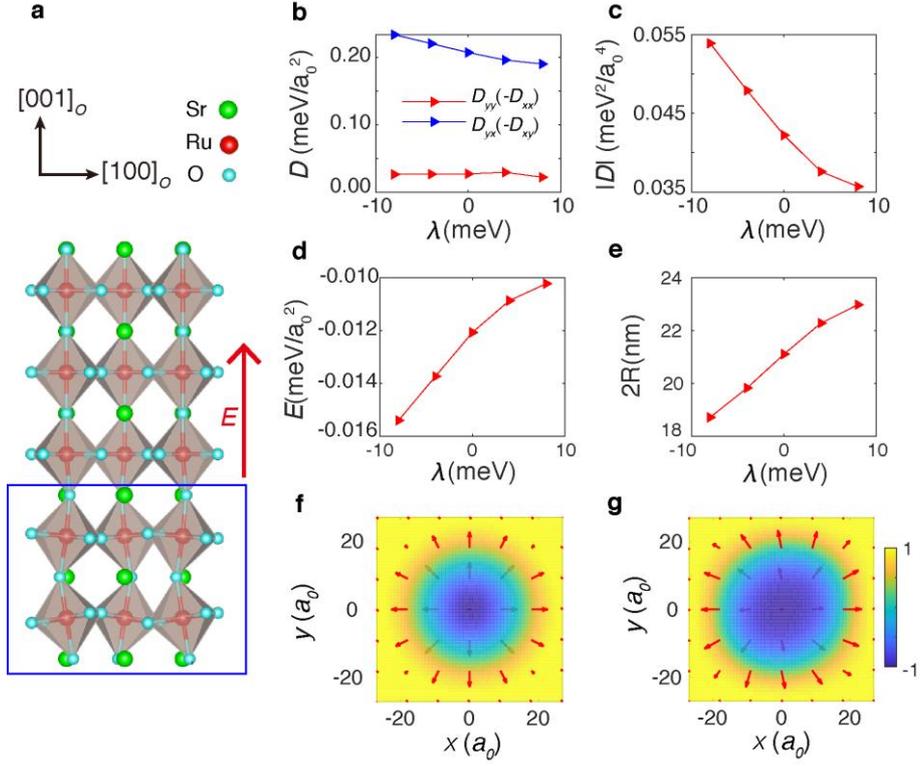

**Fig. 5 | Calculated DM interaction and skyrmion configuration. a**, Schematic view of structure model of SrRuO$_3$ slab used in the calculations. **b**, Electric-field dependence of DM interaction coefficients for the 5-u.c.-thick SrRuO$_3$ slab. $\lambda$ represents the strength of Rashba interaction, which is approximately proportional to the strength of electric-field magnitude. **c**, Electric-field dependence of determinant $|D| = D_{xx}D_{yy} - D_{xy}D_{yx}$ for DM interaction. **d**, Minimized energy density under different electric-field strengths. **e**, Skyrmion diameter determined by minimizing energy density. **f, g**, Calculated structures of skyrmion with the general form and with fully relaxed in a square grid. The arrows indicate the in-plane components of the spins, and the colors represent the out-of-plane component.